# Parallel Matrix Multiplication Using Voltage Controlled Magnetic Anisotropy Domain Wall Logic

Nicholas Zogbi, Samuel Liu, Christopher H. Bennett, Sapan Agarwal, Matthew J. Marinella, Jean Anne C. Incorvia, and T. Patrick Xiao

*Abstract*—The domain wall-magnetic tunnel junction (DW-MTJ) is a versatile device that can simultaneously store data and perform computations. These three-terminal devices are promising for digital logic due to their nonvolatility, low-energy operation, and radiation hardness. Here, we augment the DW-MTJ logic gate with voltage controlled magnetic anisotropy (VCMA) to improve the reliability of logical concatenation in the presence of realistic process variations. VCMA creates potential wells that allow for reliable and repeatable localization of domain walls. The DW-MTJ logic gate supports different fanouts, allowing for multiple inputs and outputs for a single device without affecting area. We simulate a systolic array of DW-MTJ Multiply-Accumulate (MAC) with 4-bit and 8-bit precision, which uses the nonvolatility of DW-MTJ logic gates to enable fine-grained pipelining and high parallelism. The DW-MTJ systolic array provides comparable throughput and efficiency to state-of-the-art CMOS systolic arrays while being radiation-hard. These results improve the feasibility of using domain wall-based processors, especially for extreme-environment applications such as space.

*Index Terms*—Domain Wall, Magnetic Tunnel Junction, VCMA, Logic, In-Memory Computing, Magnetism, Spintronics

## I. Introduction

THERE is a pressing need for new computational methods arising from the demand for faster and more efficient processing of data, which is currently hindered by the bottlenecks of the von Neumann architecture. Even with continuous improvements in transistor technology, processors are fundamentally limited by the speed and energy costs of accessing data, leading to an energy-inefficient architecture [1]. Using non-volatile memory as logic devices presents new opportunities to improve digital computation, potentially enabling low-voltage operation, near-zero static power dissipation, and more reliable operation compared to CMOS logic. Efficiency and reliability are important for deploying data- or processing-intensive workloads, such as artificial intelligence and machine learning algorithms, to edge or remote computing systems that have a limited power budget.

Spintronic devices provide non-volatile data storage, low read and write energy, high write endurance, and back-end-of-the-line compatibility with a CMOS fabrication process [2]. As a result of these properties, spintronics has emerged as a unique platform for dense data storage, digital and analog in-memory computing, neuromorphic computing, normally-off computing, and new, efficient implementations of digital logic [3-9]. On the reliability side, non-volatile logic gates enable rapid recovery from transient power losses. Additionally, because spintronic devices encode information in collective magnetization states rather than discrete charge, they are more resilient to the effects of ionizing and non-ionizing radiation. These advantages make spintronic computing particularly attractive for edge computing in space and defense applications [10-13].

Many spintronic architectures have been proposed for accelerating digital logic operations. These include logic circuits that use a mixture of CMOS and magnetic tunnel junction (MTJ) devices, spin wave majority gates, and devices based on magnetic domain walls (DWs) [14-21]. A practical magnetics-based digital processor for edge computing should be: (1) all-spintronic, with no accessory CMOS inside each logic gate, (2) all-electrical, requiring no external magnetic fields or optical excitation to read and write, (3) cascadable to form large circuits, having current-in/current-out or voltage-in/voltage-out operation without needing to convert data between logic gates.

The three-terminal domain wall-magnetic tunnel junction (DW-MTJ) logic gate fulfills these key requirements, while compactly implementing each logic gate (e.g. inverter, NAND) within a single nano-device [20]. In this device, a logical bit is encoded in the position of a DW along a ferromagnetic track, which forms the free layer of an MTJ. The logic state is read out and transmitted to the next logic gate through the MTJ. In our previous work, DW-MTJ prototypes have been experimentally shown to operate with both spin-transfer torque (STT) and spin-orbit torque (SOT) current input, to function as logical inverters with fanout > 1, to have electrically controllable operation with < 10% cycling variation, and to be cascadable to build circuits [22, 23]. More complex DW-MTJ circuits were benchmarked using micromagnetic simulations and compact device models [20, 24, 25]. Nevertheless, challenges remain for the reliability of DW-MTJ logic operations. Prior modeling has relied on DW inertia to sustain

This work was supported by the Laboratory Directed Research and Development Program at Sandia National Laboratories, a multimission laboratory managed and operated by National Technology & Engineering Solutions of Sandia, LLC, a wholly owned subsidiary of Honeywell International Inc., for the U.S. Department of Energy's National Nuclear Security Administration under contract DE-NA0003525. This work was also supported by the National Science Foundation Graduate Research Fellowship Program under Grant No. 2021311125 (S. Liu).

N. Zogbi, S. Liu, and J. A. C. Incorvia are with the Department of Electrical and Computer Engineering, University of Texas at Austin, Austin, TX 78712 USA (e-mail: incorvia@austin.utexas.edu).

M. J. Marinella is with Arizona State University, Tempe, AZ 85287 USA.

C. H. Bennett, S. Agarwal, and T. P. Xiao are with Sandia National Laboratories, Albuquerque, NM 87123 USA (email: txiao@sandia.gov).

the DW motion across a perfectly smooth magnetic track after the cessation of an applied current. However, in realistic tracks fabricated in scaled process nodes, this inertial motion can be rapidly halted by edge roughness or local material defects. Furthermore, the energy and performance of DW-MTJ logic has not been evaluated for modern edge computing workloads such as machine learning.

In this paper, we demonstrate that voltage-controlled magnetic anisotropy (VCMA) ensures reliable DW-MTJ logic concatenation along magnetic tracks with realistic roughness. We show using micromagnetic simulations that VCMA can electrically pin the DW at set locations, enabling deterministic and robust switching. We also demonstrate how to implement different logic functions and fanouts with minimal changes to device geometry. These device-level results are used to benchmark the energy and performance of a systolic array of DW-MTJ Multiply-Accumulate (MAC) units, a highly parallel spatial architecture that computes matrix multiplications. The non-volatility of DW-MTJ logic enables pipelined processing inside the MAC units, greatly increasing parallelism and enabling the processing throughput of DW-MTJ systolic array to approach that of state-of-the-art CMOS systolic arrays, despite the much slower switching speed of a DW-MTJ logic gate. We discuss the scaling requirements on spintronic devices and material properties to enable reliable operation and energy-efficient extreme edge computing.

## II. DW-MTJ DEVICE DESIGN

The structure and operation of the three-terminal DW-MTJ device is shown in Fig. 1. The DW is the transition region between two oppositely magnetized domains in a ferromagnetic wire, part of which also forms the free layer of an overlying MTJ. The '1' and '0' states are encoded by the MTJ resistance, which depends on the position of the DW and the MTJ's fixed layer orientation. In Fig. 1(a), the DW is on the left and the MTJ is in a high resistance state, resulting in a low output current when a voltage is applied from CLK to OUT to read the device, interpreted as a '0'. The two resistance states of the MTJ are anti-parallel (AP) $R_{AP}$ and parallel (P) $R_P$, and the tunnel magnetoresistance is defined as TMR = $(R_{AP} - R_P) / R_P$.

To write the device, current is pulsed through the IN terminal while CLK is grounded. If the current is sufficiently large, the DW moves in the same direction as the electron flow, to the right side of the device as shown in Fig. 1(b). This results in a high output current when a voltage is applied from CLK to OUT, interpreted as a '1'. The DW moves due to a combination of the spin transfer torque (STT) exerted by the spin-polarized current through the magnetic track and by the spin orbit torque (SOT) caused by current flow in the underlying heavy metal layer. The SOT effect is dominant due to the lower resistivity of the heavy metal which contains most of the current.

DW-MTJ logic gates are concatenated by connecting the OUT terminal of one device to the IN terminal of another as shown in Fig. 1(d). To transmit a stored bit from Device 1 to Device 2, a voltage pulse is applied to the CLK terminal of Device 1 and current flows through its OUT terminal. The magnitude of this current depends on Device 1's DW position. If the current exceeds Device 2's threshold, its DW moves from left to right. Concurrently, current also flows from the CLK to the IN terminal of Device 1 to reset its DW to the left, causing a current to flow from the OUT to CLK terminal of Device 0. This voltage pulse is therefore called the read-reset (RR) pulse. The current does not affect the magnetic state of Device 0. The RR scheme is desirable because by not being required to hold a state, the logic gate can be immediately re-used to process another operation [20, 22]. This enables fine-grained pipelining and parallel computation as will be described in Section VI.

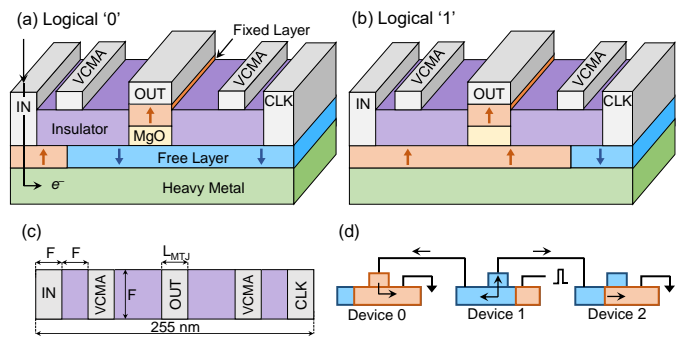

**Figure 1: DW-MTJ logical buffer. (a)** Side-view of the device when the DW is on the left and the MTJ is in a high resistance state. Orange/blue are oppositely magnetized regions. Electron flow direction during a write is indicated. **(b)** After a write, the DW is on the right and the MTJ is in a low resistance state. **(c)** Top-down view of the logic device with varying fanouts. For a fanout of 0.5, the MTJ length $L_{MTJ}$ is set to the minimum feature size F = 15 nm. For a fanout of 1, $L_{MTJ}$ is set to 3F = 45 nm. For a fanout of 2, $L_{MTJ}$ is set to 9F = 135 nm. **(d)** Side-view of a 1×3 chain of DW-MTJ buffers. Device 1 is being read-reset and Device 2 is being written.

The current that is transmitted to Device 2 corresponds to the output bit that was stored in Device 1 until the moment that the DW passes under the MTJ. After this point, Device 1 has been reset and its output current may no longer transmit the correct information. Earlier versions of the DW-MTJ gate used a very short current pulse to transmit the output bit to Device 2 before Device 1 was reset and relied on DW inertia to settle the DWs in both devices to their correct positions after the end of the current pulse [24, 25]. On a smooth ferromagnetic track, a DW can be modeled as a massive particle where it continues to move after current is turned off, as shown by micromagnetic simulations [26]. However, in a realistic magnetic track, DWs can be easily pinned by local defects or edge roughness, bringing the inertial motion to a premature halt, and causing bit errors. To overcome this challenge and make DW-MTJ logic more feasible for implementation, VCMA terminals, labeled as VCMA, are introduced in Fig. 1(a-c) to pull the DW to one of two locations on either side of the OUT terminal. To ensure non-volatility, the VCMA voltage is applied whenever the RR pulse is not applied to the device, as discussed in Section III.

Variable gate fanout is important for implementing practical logic circuits. For all fanouts and logic gate types (buffer, inverter, NAND), we use a fixed magnetic wire width equal to the minimum feature size (F = 15 nm) and a wire length of 17F = 255 nm. This keeps the threshold current and device footprint constant for all devices. The only parameter that is varied with fanout is the length of the MTJ ($L_{MTJ}$) shown in Fig. 1(c), which is set to F for fanout of 0.5, 3F for a fanout of 1, and 9F for fanout of 2. A fanout of 0.5 means that the output goes to one of the two input ports of a logic gate such as NAND, thus requiring less current per device. A fanout of 2 means one device is concatenated to two devices at its output, thus requiring greater current flow to supply data to both devices.

TABLE I
PARAMETERS USED IN THE MODEL

| Parameter | Value |
|---|---|
| Gilbert damping $\alpha$ | 0.05 |
| Saturation magnetization $M_S$ | $8 \times 10^5$ A/m |
| Exchange stiffness $A_{ex}$ | $1.3 \times 10^{-11}$ J/m |
| Uniaxial anisotropy constant $K_S(0)$ | $5 \times 10^5$ J/m$^3$ |
| Spin polarization $P$ | 0.7 |
| Temperature $T$ | 0 K |
| VCMA coefficient $\xi$ | 10 pJ/Vm |
| Clock voltage $V_{CLK}$ | 0.04 V |
| Clock pulse time $t_{RR}$ | 2 ns |
| Device rest time $t_{rest}$ | 4 ns |
| Track length $L_{wire}$ | 255 nm |
| Track width $W$ | 15 nm |
| Ferromagnet thickness $t_{FL}$ | 3 nm |
| Ferromagnet resistivity $\rho_{FL}$ | 500 μΩ·cm |
| Heavy metal thickness $t_{HM}$ | 7 nm |
| Heavy metal resistivity $\rho_{HM}$ | 40 μΩ·cm |
| Insulator dielectric constant $k_{INS}$ | 7 |
| Dielectric thickness $t_D$ | 20 nm |
| MTJ RA product | 0.675 Ω*μm$^2$ |
| MTJ parallel resistance (Fanout (FO) 0.5) $R_P$ | 3 kΩ |
| MTJ length $L_{MTJ}$ | FO 0.5: $L_{MTJ}$ = 15 nm |
| | FO 1: $L_{MTJ}$ = 45 nm |
| | FO 2: $L_{MTJ}$ = 135 nm |
| MTJ width $W_{MTJ}$ | 15 nm |

TABLE II
BUFFER CONFIGURATIONS & ENERGIES

| Initial DW1 | Initial MTJ0 | Initial MTJ1 | Reset Energy (fJ) |
|---|---|---|---|
| Right | P | P | 2.2 |
| Right | AP | P | 1.9 |
| Left | P | AP | 1.9 |
| Left | AP | AP | 1.6 |

Table I shows the parameters used for micromagnetic modeling of the device. Edge roughness is modeled by randomly removing ferromagnetic material from the edges of the track at a granularity of 1 nm, which is close to the experimentally observed length scale of roughness in ferromagnetic nanowires [27]. Random anisotropy variations in the magnetic grains within the track were also modeled.

### III. VCMA-BASED DW-MTJ CONCATENATION

The VCMA effect changes the magnetic anisotropy of the ferromagnetic wire due to an applied electric field at the insulator-CoFeB interface. This field changes the occupancy of the 3$d$ orbitals in Fe [28], which modifies the perpendicular magnetic anisotropy (PMA) of the ferromagnetic free layer. An applied voltage $V_B$ on the VCMA contacts in Fig. 1(a) reduces the PMA under these contacts, creating potential wells that confine the DW and keep the state of the logic gate stable. The difference between the local minimum in PMA and the largest PMA in the track is denoted as $\Delta K_u$. A larger $V_B$ increases $\Delta K_u$. Details of the calculation of the PMA change due to VCMA can be found in Section I of the SI.

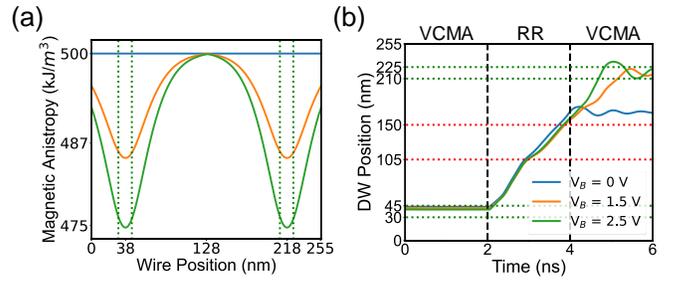

**Figure 2:** (a) Distribution of perpendicular magnetic anisotropy along the ferromagnetic free layer with three different voltages applied to the VCMA terminals. The collection of charges on the edges of the ferromagnet are not displayed. (b) Simulated DW position vs time for three voltage values when the TMR of the simulated logic devices is 115%.

Fig. 2(a) shows the effect of VCMA with three different values of $V_B$ applied to the VCMA terminals. The resulting $\Delta K_u$ are 0 kJ/m$^3$, 15 kJ/m$^3$, and 25 kJ/m$^3$ for $V_B$ of 0 V, 1.5 V, and 2.5 V respectively. These $\Delta K_u$ values show the energy wells that are formed whenever there is an applied VCMA voltage to the ferromagnetic track. In addition to this, Fig. 2(b) shows that the induced energy wells can attract the DW towards the VCMA terminals. The vertical dashed lines in Fig. 2(b) show the sequential application of VCMA, RR, and VCMA voltage pulses which modulate the DW position. The horizontal dashed lines show the lateral extent of the MTJ (105 – 150 nm) and the VCMA contacts (30 – 45 nm and 210 – 225 nm).

When there is no attractive potential for the DW ($V_B = 0$ V), the final position of the DW can be highly variable due to the stochastic effects of edge roughness, thermal noise, and timing imprecision in the RR pulse. If the DW is insufficiently driven by the RR pulse so that its final position is under or at the edges of the MTJ, the wrong logical bit can be transmitted to the next device. To reliably move the DW far to the right of the MTJ, VCMA is applied whenever RR is not applied. Stabilization of the DW to the correct position is ensured so long as the RR pulse moves it past the midpoint of the MTJ, so that it is attracted to the correct potential well as shown in Fig. 2(a). This makes the concatenation of logic more robust to process variations and noise.

### IV. SINGLE INPUT/OUTPUT LOGIC

Reliable logical concatenation of DW-MTJ devices is verified using the micromagnetic simulation software MuMax3 with parameters from Table I [29]. The circuit shown in Fig. 1(d) is used to demonstrate the concatenation of devices with a single input, a single output, and fanout of 1. The Thevenin resistance of the three concatenated DW-MTJ devices is calculated using a SPICE simulation, which is then used to compute the current density through the logic device. The amplitude of the RR pulse ($V_{CLK}$) was chosen so that DWs are propagated at close to their threshold current density, which is the minimum current density that reliably propagates the DW past the midpoint of the MTJ using our RR/VCMA scheme. For the parameters in Table I, the threshold current and threshold current density are simulated to be 2.1 μA and $7 \times 10^{10}$ A/m$^2$,

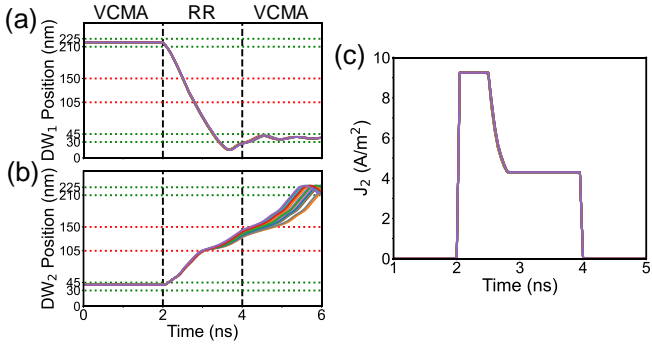

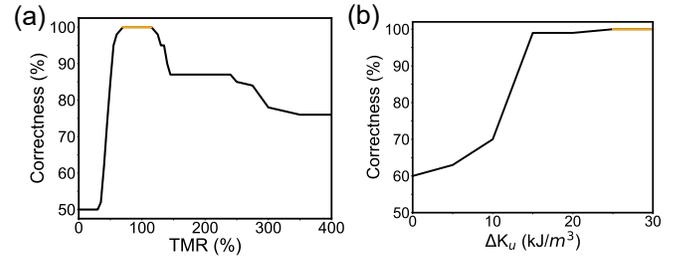

**Figure 3:** Simulation of the three-device circuit in Fig. 1(d) with TMR = 115%. DW position vs. time of (a) Device 1 during its reset pulse and (b) simultaneous DW position vs. time of Device 2. (c) Output current density from Device 1 being pulsed to Device 2.

**Figure 4:** (a) Correctness for all 8 possible configurations of the 3-device circuit vs. TMR. Yellow highlights the range with 100% correctness. (b) Correctness vs. $\Delta K_u$ when the TMR = 115%.

respectively, which are consistent with experiments on SOT-driven DW motion in CoFeB with PMA [23].

Fig. 3(a) shows the DW position vs. time for Device 1 when Device 1 is being reset and its state is transmitted to Device 2, for the three-device circuit in Fig. 1(d). The colored curves on the plot show 25 independent simulations with random grain anisotropy and varied edge roughness on the ferromagnetic track to emulate device-to-device variations. Fig. 3(b) shows the DW position in Device 2 during the same voltage pulse. Device 2's DW is driven by the current that goes through the OUT of Device 1 into the IN of Device 2 (CLK grounded), which is shown in Fig. 3(c). This result shows the successful transfer of a '1' bit from Device 1 to Device 2, which is robust to device-to-device variations. Section II in the SI shows simulation results at 300 K, which additionally show the logic gate's robustness to thermal noise.

Fig. 1(d) represents one of eight possible initial configurations for a single-input gate with a fanout of 1. There are two possible starting positions for Device 1's DW; two fixed layer orientations for Device 1, which decides whether the device acts as an inverter or a buffer; and two MTJ resistances for Device 0, which can affect the current through Device 1. Table II shows the reset energy due to the RR pulse for the four configurations where Device 1 is a buffer. Due to differences in the dynamics of the DW position and the Thevenin resistance of the circuit, some initial configurations can be more prone to errors compared to others. Supplementary Fig S4 shows that the other seven configurations possible with a fanout of 1 are also robust to device-to-device variations.

Fig. 4(a-b) shows the functionality of the logic devices by testing the correctness with 25 random simulations of each of the eight configurations of the buffer/inverter, for a total of 200 tests per data point. Each device was simulated with random variations in edge roughness and grain anisotropy on the ferromagnetic track, with a maximum anisotropy variation of 7.5 kJ/m³. The correctness is the fraction of the 200 tests that yielded the correct logical result. Fig. 4(a) shows that correct circuit operation strongly depends on TMR. When the TMR of the MTJs is less than 55%, there is a steep drop-off in correctness: the difference between $R_P$ and $R_{AP}$ is too small, which reduces the difference between a 'high' and a 'low' current seen by Device 2. The reduced difference makes it more likely that the DW in Device 2 will erroneously move in response to a 'low' current.

There is a TMR range of 75% - 115% where there are no errors. Surprisingly, for TMR > 115% there is a reduction in correctness over the eight configurations, which can be explained as follows. When Device 1 transmits a '1' to Device 2, the 2 ns pulse that is sent to Device 2 has a portion that has high current (Device 1 in P state) and low current (Device 1 in AP state), as shown in Fig. 3(c). In general, both portions contribute to the DW motion in that device, though the DW velocity is faster during the high-current portion, as shown in Fig. 3(b). With high TMR, the MTJ draws very little current in its AP state. As a result, the DW is moved very little by the low-current portion of the pulse and may fail to pass the midpoint of the MTJ. This can cause a logical '1' in Device 1 to be incorrectly passed to Device 2 as a '0'. This affects two of the eight logical configurations tested in Fig. 4, leading to the expected 75% correctness at higher TMR. While high TMR is often considered ideal for MTJs used as memory devices, here an optimal TMR is a function of the circuit design and cannot be treated as an independent figure of merit. The range of TMR for 100% correctness is wide enough to be achievable with today's MTJ processes.

The $\Delta K_u$ that is induced by VCMA also affects the logic function correctness even when a TMR of 115% is set to ensure optimal correctness, shown in Fig. 4(b). The logic is 100% correct for $\Delta K_u > 25 - 30$ kJ/m³, which corresponds to $V_B = 2.5 - 3$ V. These results show that adding VCMA to the DW-MTJ devices results in reliable and robust concatenation into circuits, and that both TMR and VCMA can be optimized to ensure 100% correctness of the circuit operation. At 300 K, the VCMA coefficient $\xi$ is reduced by 25%. To keep the same functionality, $V_B$ is increased to 3.25 - 4 V to achieve the same $\Delta K_u$ of 25 - 30 kJ/m³ [30].

## V. MULTIPLE INPUT/OUTPUT LOGIC

To obtain logic gates with different fanouts, we modify only the length of the output MTJ, which modulates the MTJ resistance. This avoids the need to adjust the supply voltage, the DW threshold current, or the layout area of logic gates with different fanout. Table III shows the reset energy for three different fanouts (0.5, 1, and 2) with a range due to different initial device configurations.

TABLE III
FANOUT ENERGIES

| Fanout | Reset Energy (fJ) |
|---|---|
| 0.5 | 1.2-1.8 |
| 1 | 1.6-2.2 |
| 2 | 2.4-3.6 |

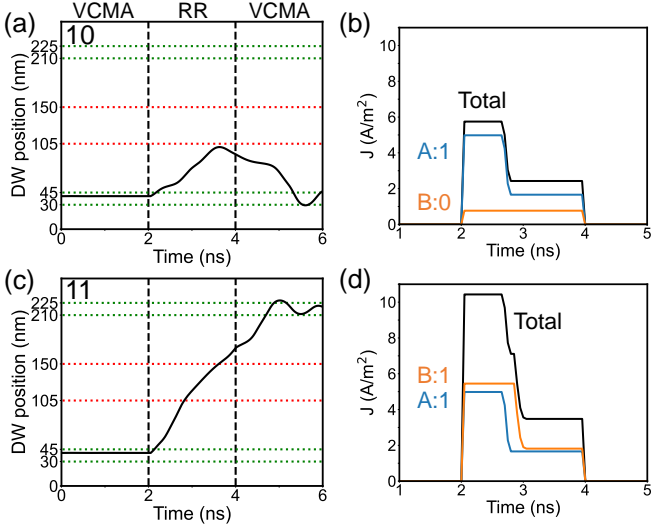

**Figure 5:** (a) Simulated DW position vs time for an AND/NAND device when the input currents are '10' and the TMR of the logic devices is 115%. (b) Input currents going into an AND/NAND device showing a logical '1' input current (blue), logical '0' input current (orange), and the total current being pulsed into the device (black). (c) Simulated DW position vs time for an AND/NAND device when the input currents are '11'. (d) Input currents going into an AND/NAND device showing the two separate logical '1' input currents (blue, orange) and the total current (black).

The logical function of a two-input logic gate depends on whether the two input devices have a fanout of 1 or 0.5. If each input has a fanout of 1, a logical OR/NOR is realized because only one input current needs to be high to propagate the DW. If each input device has a fanout of 0.5, a logical AND/NAND is realized because both currents must be high to move the DW. Schematics of these logic devices are shown in Fig. S5 in the SI. Fig. 5(a-b) shows an AND/NAND gate when the two buffer devices connected to its IN terminal are '1' and '0', which fails to propagate the DW to the opposite side. If both inputs had a logical '1', their combined current is large enough to propagate the DW to the other side of the track, as shown in Fig. 5(c-d).

To obtain a fanout of 2, the MTJ length is increased to double the output current. Calculations showing how this is achieved, accounting for series resistances in the circuit, are shown in Section IV of the SI. To ensure a fixed logic gate area and reliable concatenation of the DW-MTJ gates, we do not further increase the MTJ length to increase fanout. Instead, we use a cascade of fanout-2 buffers to obtain larger fanouts. More details on design choices related to fanout can be found in Section IV of the SI.

## VI. PIPELINED DW-MTJ MATRIX MULTIPLICATION

The dual functionality of a DW-MTJ as a logic gate and as a non-volatile memory element allows the device to efficiently implement a form of sequential logic that greatly increases data parallelism compared to the conventional combinatorial logic used in CMOS processors. Fig. 6 shows how DW-MTJ logic gates are connected to form a full adder. Because data is passed between gates only during clock pulses, buffers are used to keep the data aligned in time. A logic gate cannot simultaneously receive and transmit data, To avoid interference, it also cannot receive data while its output gate is being reset. Therefore, at any moment, each gate operates in one of three modes: receive, transmit, or standby. Gates that are aligned in their logical depth operate in the same mode, shown by the colored bands in Fig. 6. A three-phase clock ($\Phi_1$, $\Phi_2$, and $\Phi_3$) switches each band of gates from one operational mode to the next. Each mode lasts for one phase of the clock period (4 ns). All gates cycle through all three modes within a full clock period (12 ns).

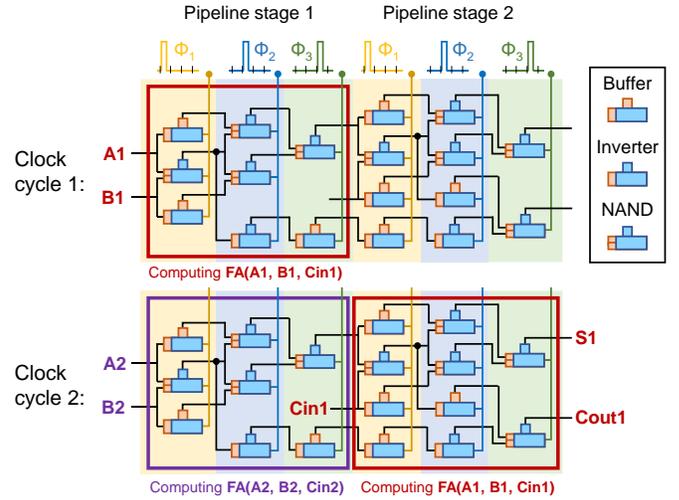

**Figure 6:** Illustration of pipelined logic in a DW-MTJ full adder. Data moves through the circuit one logical stage at a time, and gates in the same stage are driven by the same signal from a three-phase clock. Data is pipelined through each set of three stages, which takes one clock cycle to traverse. DW-MTJs serve both as logic gates and as storage elements for pipelining.

In the transmit mode, an RR pulse ($t_{RR}$ = 2 ns) is applied to the CLK terminal to reset the device and transmit its state to the next device, then a VCMA pulse ($t_{rest}$ = 2 ns) is applied to pin the DW on the left side. In the receive mode, the DW is driven by the RR pulse applied to its input device, then a VCMA pulse is applied to pin the DW on the left or right. In the standby mode, the DW is kept stationary for 4 ns using VCMA.

Each DW-MTJ logic gate receives and processes a new set of inputs on every clock cycle. This allows a single logic circuit to concurrently process many independent operations that are pipelined through the stages of computation on each clock cycle. The data buffering that is needed for pipelining is implemented by the DW-MTJ logic gates themselves, and the size of the pipeline stage is a window of logical depth three as shown in Fig. 6. This fine-grained pipelining allows concurrent computation on much more data than standard instruction-level pipelining. It enables high throughput while only fitting a few sequential operations into a clock cycle, which compensates for the relatively slow switching speed of a DW-MTJ gate compared to CMOS. Pipelining at this granularity is possible in CMOS but is impractical due to the area and power consumption of the many pipeline registers required [31, 32]. However, it comes at no cost in a DW-MTJ processor.

A disadvantage of fine-grained pipelining is that the cost of flushing the pipeline upon an interrupt is high, because a large amount of data is stored by the DW-MTJ gates in the pipeline. This makes the logic scheme a poor fit for general-purpose computing where pipeline flushes are common (e.g. due to branch mispredictions). However, fine-grained pipelining is beneficial for processing data in large blocks, where pipeline flushes are extremely rare. As a notable example, DW-MTJ logic is well suited to specialized accelerators for matrix vector multiplication (MVM), which are growing in importance due to the growth of machine learning workloads [33]. The building block of these accelerators is the MAC operation: D=A×B+C. We focus on MACs where A, B, C, and D are multi-bit integers. Integer MACs are the main computational primitive for deep neural network inference, with a typical resolution of 8 bits [34], though inference at lower precision (e.g. 4 bits) is popular for low-power applications [35].

Fig. 7(a) shows a 4-bit DW-MTJ MAC unit. The circuit consists of a 4-bit multiplier to compute A×B, and the product is passed to an 8-bit ripple carry adder which adds this result to the operand C. An array multiplier is used, with the same logical structure as a CMOS implementation. To enable fine-grained pipelining, DW-MTJ buffers are inserted into the circuit as needed to ensure that the operands arrive at the correct gates at the correct times. We choose a pipelined array multiplier, rather than a bit serial multiplier, to ensure that a full MAC is completed on every clock cycle. In a given cycle, one bit of one MAC output is available at every output bit position (e.g. D8 of MAC 1, D7 of MAC 2, ..., D0 of MAC 9). This format allows the output terminal "D" of one MAC to be directly connected to the input terminal "C" of another MAC. The 4-bit multiplier has 13 pipeline stages and processes 13 MACs in parallel. Data is buffered by one full clock cycle using a buffer chain, which is a linear chain of three buffers or a tree of buffers with depth three, depending on the needed fanout (1 to 8). The AND chain is the same as the buffer chain but has an AND gate as a head.

Fig. 7(b) and (c) show how MAC units can be concatenated to process MVMs: $\mathbf{W}\vec{x}$. To maximize parallelism, we use a systolic array architecture to process neural network inference, where the matrix $\mathbf{W}$ is usually fixed [36]. The MAC unit stores the fixed operand, which is an element of $\mathbf{W}$. Inputs are broadcast horizontally across the array while partial sums are accumulated vertically, and processing is pipelined at the MAC level. In the DW-MTJ systolic array, we leverage a second level of fine-grained pipelining within the MAC units themselves. This greatly increases the MVM throughput over a DW-MTJ processor without fine-grained pipelining.

The DW-MTJ pipelined systolic array operation is shown in Fig. 7(c), illustrated using 4-bit MACs. The first element of the vector, $x_1$ (4 bits), arrives on the top left MAC unit which stores $W_{11}$ (4 bits). The product $W_{11}x_1$ (8 bits) is passed downward to the "C" input port of the next MAC unit. This unit computes $W_{21}x_2$, with a 2-cycle delay relative to the first MAC unit, then performs the addition $W_{11}x_1 + W_{21}x_2$ (9 bits). Fine-grained pipelining allows this addition to begin on the less significant bits while the more significant bits of $W_{11}x_1$ and $W_{21}x_2$ are still being computed by their respective multipliers. This partial sum is again passed downward, so that the output of the bottom MAC unit on the column is the dot product $\sum W_{i1}x_i$. To support a sum of 256 elements, we extend the ripple carry adder inside each 4-bit MAC unit to 16 bits. For a 256×256 systolic array with 8-bit MACs, we extend the array multiplier to a precision of 8 bits and the ripple carry adder to a precision of 24 bits. Each MAC unit also passes the input $x_i$ to the right. Thus, each column computes an independent dot product, with a 1-cycle delay between columns, to implement an MVM.

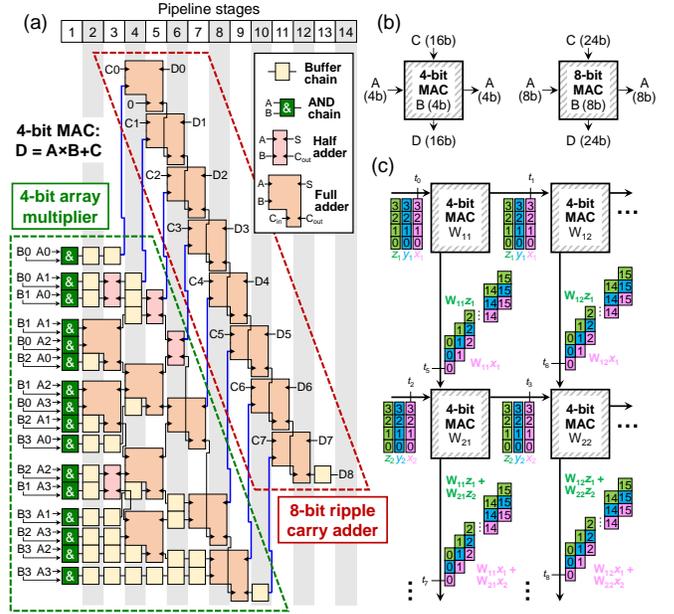

**Figure 7:** (a) DW-MTJ MAC unit with 4-bit multiply, 8-bit accumulate. (b) Symbol for a 4-bit MAC and 8-bit MAC unit. The 4-bit MAC unit has a 16-bit adder and the 8-bit MAC has a 24-bit adder to support the accumulation of 256 products. (c) Systolic array processing of MVM using 4-bit MAC units.

## VII. PERFORMANCE, ENERGY, AND AREA

We evaluate a 256×256, 8-bit DW-MTJ systolic array to match the size and precision of the systolic array in the Google TPUv1 [37]. We also evaluate a 4-bit systolic array of the same size for low-energy inference applications. Both systolic arrays complete a new MVM on every clock cycle.

The peak throughput of the TPUv1 is 92 TOPS (TeraOperations/s), while that of the DW-MTJ systolic array is 10.9 TOPS when operated at 0K using the parameters in Table I. Operation at 300 K reduces the threshold current density for DW motion and increases the DW susceptibility, which enables a reduction in both the amplitude and length of the RR pulse. This improves the throughput to 14.5 TOPS: see Section III in SI for details. Though the DW-MTJ throughput is lower by ~6×, fine-grained pipelining allows the DW-MTJ system to make up for a much larger speed deficit at the logic gate level. The DW-MTJ gate delay (4 ns) is ~4000× larger than that of a CMOS logic gate (~1 ps) at a similar 15 nm process node [38].

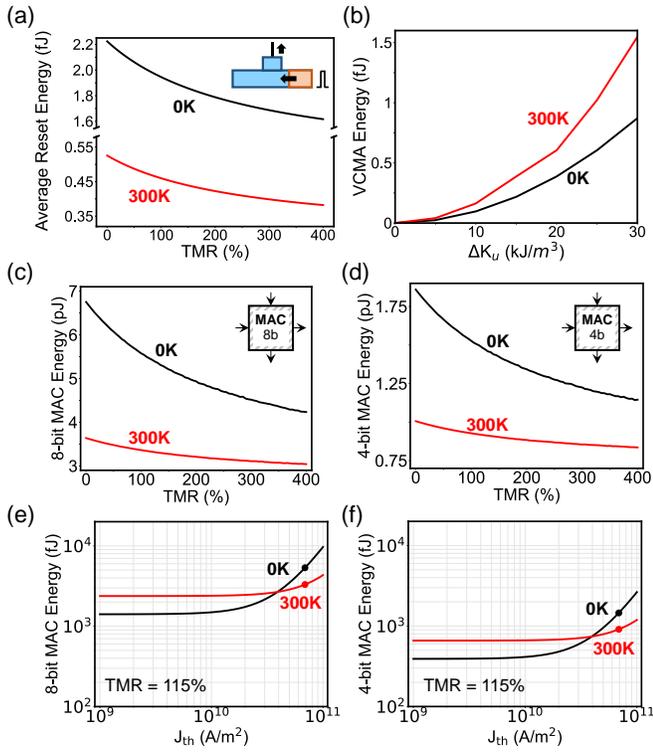

**Figure 8:** (a) Reset energy vs. TMR for a DW-MTJ buffer, averaged over possible DW-MTJ initial states. (b) Energy to apply a VCMA voltage vs. PMA change per logic gate. (c) Energy per 8-bit MAC vs. TMR. Each point is averaged over 100 random test MACs. (d) Energy per 4-bit MAC vs TMR. (e) Projected energy per 8-bit MAC vs DW threshold current density at TMR = 115%. The dot marks the $J_{th}$ value derived from the parameters in Table I. (f) Projected energy per 4-bit MAC vs $J_{th}$.

The energy to concatenate two logic gates is the sum of the energy to propagate the DWs using the RR pulse and the energy to stabilize the DWs using VCMA. The former, called the reset energy, is shown in Fig. 8(a). Higher TMR increases the MTJ's AP resistance, which reduces the current and the energy. A TMR of 115% yields the optimal correctness for the assumed device parameters. Fig. 8(b) shows the energy to supply the VCMA pulse vs. the depth of the PMA potential well, $\Delta K_u$. For optimal correctness, we use $\Delta K_u = 25$ kJ/m$^3$, which is obtained at $V_B = 2.5$ V ($V_B = 3.25$ V at 300 K). The energy in Fig. 8(b) is the $CV^2$ energy of charging the VCMA capacitors (purple dielectric in Fig. 1(a)) and the metal lines used to supply the VCMA pulses from its generation circuit. For the metal line capacitance, we assumed a compact layout where many gates that are aligned in their clock phase are arranged in a column. Devices in a column share a set of clocking and VCMA lines. The gates are separated by 100 nm to avoid coupling through their stray magnetic fields. For a typical 14 nm process, this separation leads to a metal line capacitance per device of 40 aF for the two VCMA electrodes and 20 aF for the CLK electrode [39]. At a feature size of F = 15 nm, every DW-MTJ logic gate has the same footprint of ~181.5F$^2$, which includes the distance between adjacent gates to avoid lateral coupling [40]. This is ~2× smaller than a typical CMOS NAND gate [38].

We developed a simulator to model the functionality and energy of arbitrary digital logic circuits constructed from DW-MTJ gates. Circuit-level energy consumption is based on gate-level energies found from micromagnetic simulations, accounting for energy differences due to device state (P vs AP) and fanout. Fig. 8(c) shows the energy per 8-bit MAC of a 256×256 DW-MTJ systolic array, using the device parameters in Table I. At TMR = 115%, the systolic array's efficiency is 5.4 pJ/MAC, or 0.37 TOPS/W (0.87 TOPS/W at 300 K). This is comparable to state-of-the-art digital CMOS accelerators for neural network inference [33, 41]. Fig. 8(d) shows the energy per 4-bit MAC. At TMR = 115%, the energy efficiency is 1.3 TOPS/W (2.2 TOPS/W at 300K).

Fig. 8(e-f) project how the DW-MTJ systolic array's efficiency scales with the threshold current density $J_{th}$ for DW motion. A 10× reduction in $J_{th}$ relative to the simulated device improves the efficiency to 1.4 pJ/MAC (1.4 TOPS/W) and 400 fJ/MAC (5 TOPS/W) for an 8-bit MAC and 4-bit MAC, respectively, at 0K. At these low current densities for DW propagation, the efficiency is limited by the VCMA energy for DW pinning. At 300K, the efficiency in this limit is 2.4 pJ/MAC (0.83 TOPS/W) and 660 fJ/MAC (3.0 TOPS/W) for an 8-bit MAC and 4-bit MAC respectively. Reduction in the DW threshold current density below $10^{10}$ A/m$^2$ can be accomplished by optimization of the SOT device geometry and current injection mechanism, as well as through tighter control of edge roughness and defects in the thin ferromagnetic strip [42].

## VIII. Conclusion

We have demonstrated the ability to perform reliable and robust concatenation of logic using the three-terminal DW-MTJ device with the assistance of VCMA. This work shows that TMR and VCMA can be optimized to ensure 100% correctness of logic circuit operations, using values that are accessible with current MTJ fabrication processes. Different logic gate types and fanouts can be implemented with no change to the device footprint. Additionally, we evaluated DW-MTJ systolic arrays for computing MVMs with 4-bit and 8-bit MAC precision. The resulting energy per MAC operation is on par with state-of-the-art CMOS accelerators for neural network inference, while providing a high throughput (14.5 TOPS at 8-bit, 300K) despite the slower switching speed of DW-MTJ devices. These results show that non-volatile spintronic logic devices can be used to effectively accelerate edge computing applications while offering robustness under extreme environments.

## Acknowledgment

The authors acknowledge the Texas Advanced Computing Center (TACC) at The University of Texas at Austin for providing HPC resources that have contributed to the research results reported within this paper. URL: http://www.tacc.utexas.edu. The authors acknowledge support from the National Science Foundation Graduate Research Fellowship under Grant No. 2021311125 (S.L.).

This work was funded in part by the Rad Edge Grand Challenge Laboratory Directed Research & Development (LDRD) project at Sandia National Laboratories. This article has been authored by employees of National Technology &

Engineering Solutions of Sandia, LLC under Contract No. DE-NA0003525 with the U.S. Department of Energy (DOE). The employee owns all right, title and interest in and to the article and is solely responsible for its contents. The United States Government retains and the publisher, by accepting the article for publication, acknowledges that the United States Government retains a non-exclusive, paid-up, irrevocable, world-wide license to publish or reproduce the published form of this article or allow others to do so, for United States Government purposes. The DOE will provide public access to these results of federally sponsored research in accordance with the DOE Public Access Plan https://www.energy.gov/downloads/doe-public-access-plan.
## REFERENCES

[1] J. Shalf, "The future of computing beyond Moore's Law," *Philos Trans A Math Phys Eng Sci,* vol. 378, no. 2166, p. 20190061, 2020.
[2] "CMOS with a spin," *Nature Electronics,* vol. 2, no. 7, pp. 263-263, 2019/07/01 2019.
[3] P. Barla, V. K. Joshi, and S. Bhat, "Spintronic devices: a promising alternative to CMOS devices," *Journal of Computational Electronics,* vol. 20, no. 2, pp. 805-837, 2021.
[4] S. Jung *et al.*, "A crossbar array of magnetoresistive memory devices for in-memory computing," *Nature,* vol. 601, no. 7892, pp. 211-216, 2022.
[5] S. Liu, T. P. Xiao, C. Cui, J. A. C. Incorvia, C. H. Bennett, and M. J. Marinella, "A domain wall-magnetic tunnel junction artificial synapse with notched geometry for accurate and efficient training of deep neural networks," *Applied Physics Letters,* vol. 118, no. 20, 2021.
[6] S. A. Siddiqui, S. Dutta, A. Tang, L. Liu, C. A. Ross, and M. A. Baldo, "Magnetic Domain Wall Based Synaptic and Activation Function Generator for Neuromorphic Accelerators," *Nano Lett,* vol. 20, no. 2, pp. 1033-1040, 2020.
[7] S. Jain, A. Ranjan, K. Roy, and A. Raghunathan, "Computing in Memory With Spin-Transfer Torque Magnetic RAM," *IEEE Transactions on Very Large Scale Integration (VLSI) Systems,* vol. 26, no. 3, pp. 470-483, 2018.
[8] T. Leonard *et al.*, "Shape-Dependent Multi-Weight Magnetic Artificial Synapses for Neuromorphic Computing," *Advanced Electronic Materials,* vol.8, no. 2, 2022.
[9] W. Hwang *et al.*, "Energy Efficient Computing with High-Density, Field-Free STT-Assisted SOT-MRAM (SAS-MRAM)," *IEEE Transactions on Magnetics,* vol. 59, no.3, pp. 1-1, 2022.
[10] S. Gerardin and A. Paccagnella, "Present and Future Non-Volatile Memories for Space," *IEEE Transactions on Nuclear Science,* vol. 57, no. 6, pp. 3016-3039, 2010.
[11] H. Hughes *et al.*, "Radiation Studies of Spin-Transfer Torque Materials and Devices," *IEEE Transactions on Nuclear Science,* vol. 59, no. 6, pp. 3027-3033, 2012.
[12] R. Fanghui, A. Jander, P. Dhagat, and C. Nordman, "Radiation Tolerance of Magnetic Tunnel Junctions With MgO Tunnel Barriers," *IEEE Transactions on Nuclear Science,* vol. 59, no. 6, pp. 3034-3038, 2012.
[13] M. Alamdar *et al.*, "Irradiation Effects on Perpendicular Anisotropy Spin–Orbit Torque Magnetic Tunnel Junctions," *IEEE Transactions on Nuclear Science,* vol. 68, no. 5, pp. 665-670, 2021.
[14] T. Fischer *et al.*, "Experimental prototype of a spin-wave majority gate," *Applied Physics Letters,* vol. 110, no. 15, p. 152401, 2017.
[15] D. E. Nikonov, G. I. Bourianoff, and T. Ghani, "Proposal of a Spin Torque Majority Gate Logic," *IEEE Electron Device Letters,* vol. 32, no. 8, pp. 1128-1130, 2011.
[16] M. Kazemi, E. Ipek, and E. G. Friedman, "Energy-Efficient Nonvolatile Flip-Flop With Subnanosecond Data Backup Time for Fine-Grain Power Gating," *IEEE Transactions on Circuits and Systems II: Express Briefs,* vol. 62, no. 12, pp. 1154-1158, 2015.
[17] E. Raymenants *et al.*, "All-Electrical Control of Scaled Spin Logic Devices Based on Domain Wall Motion," *IEEE Transactions on Electron Devices,* vol. 68, no. 4, pp. 2116-2122, 2021.
[18] Z. Luo *et al.*, "Current-driven magnetic domain-wall logic," *Nature,* vol. 579, no. 7798, pp. 214-218, Mar 2020, doi: 10.1038/s41586-020-2061-y.
[19] D. Morris, D. Bromberg, J.-G. Zhu, and L. Pileggi, "mLogic: Ultra-Low Voltage Non-Volatile Logic Circuits Using STT-MTJ Devices," *Design Automation Conference* (DAC), 2012.
[20] J. A. Currivan, J. Youngman, M. D. Mascaro, M. A. Baldo, and C. A. Ross, "Low Energy Magnetic Domain Wall Logic in Short, Narrow, Ferromagnetic Wires," *IEEE Magnetics Letters,* vol. 3, 3000104, 2012.
[21] Z. Chowdhury *et al.*, "Efficient In-Memory Processing Using Spintronics," *IEEE Computer Architecture Letters,* vol. 17, no. 1, pp. 42-46, 2018.
[22] J. A. Currivan-Incorvia *et al.*, "Logic circuit prototypes for three-terminal magnetic tunnel junctions with mobile domain walls," *Nat Commun,* vol. 7, p. 10275, 2016.
[23] M. Alamdar *et al.*, "Domain wall-magnetic tunnel junction spin–orbit torque devices and circuits for in-memory computing," *Applied Physics Letters,* vol. 118, no. 11, 2021.
[24] T. P. Xiao *et al.*, "Energy and Performance Benchmarking of a Domain Wall-Magnetic Tunnel Junction Multibit Adder," *IEEE Journal on Exploratory Solid-State Computational Devices and Circuits,* vol. 5, no. 2, pp. 188-196, 2019.
[25] X. Hu, A. Timm, W. H. Brigner, J. A. C. Incorvia, and J. S. Friedman, "SPICE-Only Model for Spin-Transfer Torque Domain Wall MTJ Logic," *IEEE Transactions on Electron Devices,* vol. 66, no. 6, pp. 2817-2821, 2019.
[26] K. D. Belashchenko, O. Tchernyshyov, A. A. Kovalev, and O. A. Tretiakov, "Magnetoelectric domain wall dynamics and its implications for magnetoelectric memory," *Applied Physics Letters,* vol. 108, no. 13, 2016.
[27] S. Dutta, S. A. Siddiqui, J. A. Currivan-Incorvia, C. A. Ross, and M. A. Baldo, "The Spatial Resolution Limit for an Individual Domain Wall in Magnetic Nanowires," *Nano Lett,* vol. 17, no. 9, pp. 5869-5874, 2017.
[28] J. Zhang, P. V. Lukashev, S. S. Jaswal, and E. Y. Tsymbal, "Model of orbital populations for voltage-controlled magnetic anisotropy in transition-metal thin films," *Physical Review B,* vol. 96, no. 1, 2017.
[29] A. Vansteenkiste, J. Leliaert, M. Dvornik, M. Helsen, F. Garcia-Sanchez, and B. Van Waeyenberge, "The design and verification of MuMax3," *AIP Advances,* vol. 4, no. 10, 2014.
[30] A. Sugihara *et al.*, "Temperature dependence of higher-order magnetic anisotropy constants and voltage-controlled magnetic anisotropy effect in a Cr/Fe/MgO junction," *Japanese Journal of Applied Physics,* vol. 59, no. 1, 2020.
[31] F. Lu and H. Samueli, "A 200 MHz CMOS pipelined multiplier-accumulator using a quasi-domino dynamic full-adder cell design," *IEEE Journal on Solid-State Circuits,* vol. 28, no. 2, pp. 123-132, 1993.
[32] T. G. Noll, D. Schmitt-Landsiedel, H. Klar, and G. Enders, "A Pipelined 330-MHz Multiplier," *IEEE Journal of Solid-State Circuits,* vol. 21, no. 3, pp. 411-416, 1986.
[33] V. Sze, Y.-H. Chen, T.-J. Yang, and J. S. Emer, "Efficient Processing of Deep Neural Networks: A Tutorial and Survey," *Proceedings of the IEEE,* vol. 105, no. 12, pp. 2295-2329, 2017.
[34] B. Jacob *et al.*, "Quantization and Training of Neural Networks for Efficient Integer-Arithmetic-Only Inference," *IEEE/CVF Conference on Computer Vision and Pattern Recognition*, 2018.
[35] J. Choi, Z. Wang, S. Venkataramani, P. I.-J. Chuang, V. Srinivasan, and K. Gopalakrishnan, "PACT: Parameterized clipping activation for quantized neural networks," *arXiv preprint arXiv:1805.06085,* 2018.
[36] H. T. Kung, "Why Systolic Architectures?," *Computer,* vol. 5, no. 1, pp. 37-46, 1982.
[37] N. P. Jouppi *et al.*, "In-Datacenter Performance Analysis of a Tensor Processing Unit," *International Symposium on Computer Architecture* (ISCA), 2017.
[38] D. E. Nikonov and I. A. Young, "Overview of Beyond-CMOS Devices and a Uniform Methodology for Their Benchmarking," *Proceedings of the IEEE,* vol. 101, no. 12, pp. 2498-2533, 2013.
[39] M. J. Marinella *et al.*, "Multiscale Co-Design Analysis of Energy, Latency, Area, and Accuracy of a ReRAM Analog Neural Training Accelerator," *IEEE Journal on Emerging and Selected Topics in Circuits and Systems,* vol. 8, no. 1, pp. 86-101, 2018.
[40] C. Cui *et al.*, "Maximized lateral inhibition in paired magnetic domain wall racetracks for neuromorphic computing," *Nanotechnology,* vol. 31, no. 29, p. 294001, 2020.
[41] A. Reuther, P. Michaleas, M. Jones, V. Gadepally, S. Samsi, and J. Kepner, "AI Accelerator Survey and Trends," *IEEE High Performance Extreme Computing Conference* (HPEC), 2021.
[42] A. V. Khvalkovskiy *et al.*, "Matching domain-wall configuration and spin-orbit torques for efficient domain-wall motion," *Physical Review B,* vol. 87, no. 2, 2013.
*Zogbi, Liu,* et al. *Page 8 of 8*

# *Supplementary Information*: Parallel Matrix Multiplication Using Voltage Controlled Magnetic Anisotropy Domain Wall Logic

Nicholas Zogbi[1], Samuel Liu[1], Christopher H. Bennett[2], Sapan Agarwal[2], Matthew J. Marinella[3], Jean Anne C. Incorvia[1], and T. Patrick Xiao[2]

[1]*Department of Electrical and Computer Engineering, University of Texas at Austin, Austin, TX 78712*
[2]*Sandia National Laboratories, Albuquerque, NM 87123*
[3]*Department of Electrical, Computer and Energy Engineering, Arizona State University, Tempe, AZ 85281*

## I. VCMA Calculation methods

Under the voltage-controlled magnetic anisotropy (VCMA) effect, an applied electric field at the insulator-ferromagnet interface changes the perpendicular magnetic anisotropy (PMA) of the ferromagnetic free layer. The effect of the field on PMA is given by:

$$K_s(V_B) = K_s(0) - \frac{\xi E}{t_{FM}} \quad (1)$$

where $K_s(V_B)$ is the PMA at an applied voltage $V_B$, $K_s(0)$ is the PMA when there is no applied voltage to the ferromagnetic free layer, E is the electric field being applied to the DW wire, $\xi$ is the VCMA coefficient, and $t_{FM}$ is the thickness of the ferromagnetic track where all the parameters used in the calculation are shown in Table I [1, 2].

To study the effect of VCMA on the DW-MTJ device, an analytical model is constructed that calculates the effective electric field at the insulator free layer interface when a voltage $V_B$ is applied. As stated in the text, the Jacobi Method [3] was used to calculate the electric potential across the device and the gradient of this was taken in order to find the electric field. Using (1), the resulting perpendicular magnetic anisotropy when $V_B$ = 2.5 V can be expressed as a 14$^{th}$ order polynomial,

$$K_s(V) = \sum_{n=0}^{14} a_n x^n \quad (1)$$

| $a_0 = 5 \times 10^5$ | $a_1 = 4.1 \times 10^0$ | $a_2 = -2.8 \times 10^0$ | $a_3 = -2.3 \times 10^{-3}$ |
| --- | --- | --- | --- |
| $a_4 = 8.7 \times 10^{-4}$ | $a_5 = 3.5 \times 10^{-7}$ | $a_6 = -5.3 \times 10^{-7}$ | $a_7 = -9.6 \times 10^{-12}$ |
| $a_8 = 9.8 \times 10^{-11}$ | $a_9 = -1.4 \times 10^{-15}$ | $a_{10} = -7.6 \times 10^{-15}$ | $a_{11} = 1.0 \times 10^{-19}$ |
| $a_{12} = 2.7 \times 10^{-19}$ | $a_{13} = -1.9 \times 10^{-24}$ | $a_{14} = -3.6 \times 10^{-24}$ | |

where $x$ is the position on the ferromagnetic track. This polynomial was integrated with the Mumax3 script allowing for modifications to anisotropy regions on the ferromagnetic wire whenever there is no current pulsing through the device.

We note that Equation (1) and $\xi$ are valid only up to a maximum electric field of ~100 mV/nm or $\Delta K_u$ = 325 kJ/m$^3$ [4], but as we later show, VCMA can guarantee correctness at much smaller fields.

## II. Temperature Simulations

Fig. S1 shows a read-reset operation when simulated at room temperature with thermal noise with the three-device configuration in Fig. 1(d) where both device 0 and device 1 MTJs are parallel and the starting DW position is on the right side of the track. Fig. S2 shows a read-reset operation when the three-device configuration is set to Fig. S3(c). These two configurations were chosen because they are the most prone to errors, due to the amplitude of the output current density from device 1. The configuration used in Fig. S1 will have the lowest amount of current density that should propagate the DW to the opposite side for the concatenated device. The configuration used in Fig. S2 is prone to errors because it will be the highest amount of current density that will not propagate the DW to the other side for the concatenated device.

In order to conduct these simulations at room temperature, modifications to the pulse time and input voltage were required. At 300 K, DW propagation is easier because it is more susceptible to current. Therefore, the threshold current of the device is lowered, allowing for the voltage of the RR pulse to be reduced. The time of this pulse can also be reduced because the DW's velocity is greater at room temperature. Thus, the input voltage ($V_{CLK}$) and RR pulse time ($t_{RR}$) are reduced to 27.5 mV and 1 ns. Fig. S3 shows logical 1 and 0 operations when the fanout of the device is 0.5. The DW motion also becomes more susceptible to thermal noise at room temperature, but our results show that this added perturbation can be suppressed by the pinning effect of the VCMA-induced anisotropy potentials. We focused on the most error-prone configurations, rather than checking the correctness of every configuration, as there is a large computational overhead when conducting these simulations at room temperature.

## III. Temperature Performance and Energy

Due to simulations at room temperature causing parameters to be modified, the performance and energy of the systolic array will change. Since the pulse time is reduced to 1 ns at room temperature (see SI Section II), one phase of the clock period will reduce from 4 ns to 3 ns allowing for a full clock period of 9 ns. This reduction to the clock cycle allows for 14.5 TOPS for the systolic array which is a ~1.3× increase compared to the 0 K simulations. Furthermore, at room temperature, the systolic array's efficiency is reduced to 3.3 pJ per 8-bit MAC (0.6 TOPS/W) and 918 fJ per 4-bit MAC (2.2 TOPS/W). This is a ~1.6× reduction in efficiency compared to the simulations conducted at 0 K.

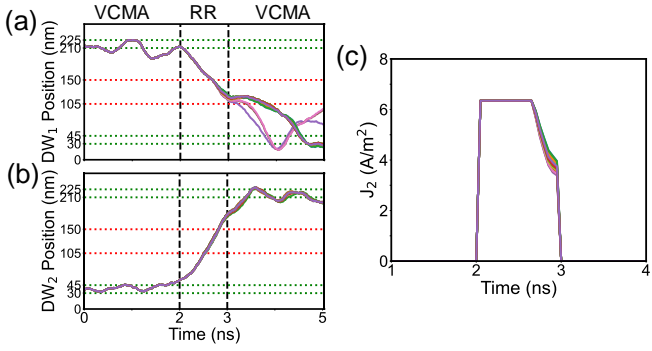

**Figure S1:** Simulation results for the three-device circuit at room temperature with TMR = 115%. DW position vs. time of (a) Device 1 during its reset pulse and (b) simultaneous DW position vs. time of Device 2. (c) Output current density from Device 1 being pulsed to Device 2.

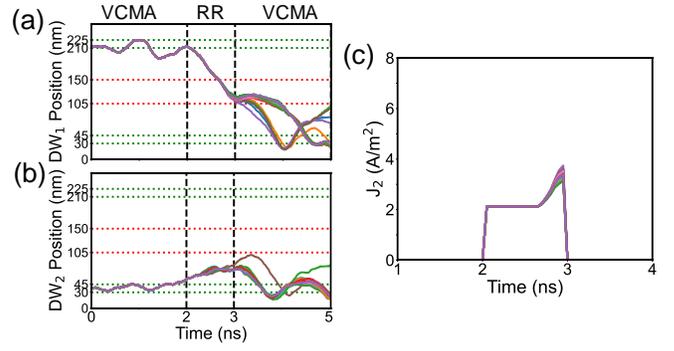

**Figure S2:** Simulation results for the three-device circuit at room temperature with TMR = 115%. DW position vs. time of (a) Device 1 during its reset pulse and (b) simultaneous DW position vs. time of Device 2. (c) Output current density from Device 1 being pulsed to Device 2.

## IV. DEVICE FANOUTS

A unique feature of the DW-MTJ logic gates is that the footprint for the devices stays constant no matter the fanout. However, the current resulting from the RR pulse needs to vary depending on the fanout of the device. To obtain this feature, the length of the MTJ is varied with lengths of 15 nm, 45 nm, and 135 nm for a fanout of 0.5, 1, and 2 respectively while the width of the device is 15 nm to obtain an output current from the RR pulse of 0.5J, J, and 2J. Thus, the current density is modeled based on the resistance and the Thevenin model into an equation expressed as

$$J = J_x \frac{R_0}{R_0 + R_1} \quad (2)$$

where J is resulting current from the RR pulse going to the concatenated device, $J_x$ is the effective applied current, $R_0$ is the effective resistance of device 0, and $R_1$ is the effective resistance of device 1. These quantities are given by the equations below.

$$J_x = V_{CLK} \frac{W(t_{FL} + t_{HM})}{R_{th}} \quad (3)$$

$$R_{0,1} = R_{MTJ_{0,1}} + \frac{\rho_{FL}\frac{L_{wire}}{W*t_{FL}} * \rho_{HM}\frac{L_{wire}}{W*t_{HM}}}{\rho_{FL}\frac{L_{wire}}{W*t_{FL}} + \rho_{HM}\frac{L_{wire}}{W*t_{HM}}} \quad (4)$$

$$R_{MTJ_{0,1}} = R_{P,MTJ_{0,1}} \text{ or } TMR* R_{P,MTJ_{0,1}} + R_{P,MTJ_{0,1}} \quad (5)$$

$$R_{P,MTJ_{0,1}} = \frac{R_P}{3^{FO}}, \text{if } FO > \frac{1}{2}, \text{else } 3 \text{ k}\Omega \quad (6)$$

All the parameters from the presented equations are shown in Table I in the main text.

To achieve large fanouts (>2), we use a cascade of buffers with a fanout of 2, as described in the main text. Fanouts >2 can be achieved by increasing the MTJ length, but the circuit design becomes more challenging since the total length of the logic devices must be increased: e.g. a fanout of 4 requires device length of 27F (1.335 μm). Additionally, if the device length is too long, VCMA could no longer be used effectively, since $V_B$ applied to the VCMA terminals would cause little to no change in PMA in the middle of the wire. For reliable control of DW position, the VCMA contacts for DW pinning must be located near the ends of the wire. Otherwise, if the VCMA contacts are too close to the central MTJ, there is a chance for the RR pulse to drive the DW to the ends of the wire and annihilate it. If a longer device were used to provide large fanout, the DW would need to propagate a longer distance between VCMA contacts. A longer device would therefore require pulses with higher amplitude or duration for reliable logic concatenation, which increases energy and adds complexity to the circuit design.

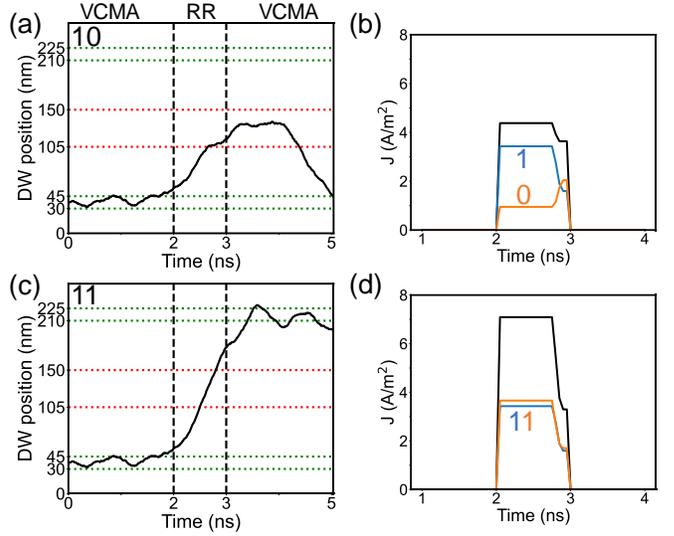

**Figure S3:** (a) Simulated DW position vs time at room temperature for an AND/NAND device when the input currents are '10' and the TMR of the logic devices is 115%. (b) Input currents going into an AND/NAND device showing a logical '1' input current (blue), logical '0' input current (orange), and the total current being pulsed into the device (black). (c) Simulated DW position vs time for an AND/NAND device when the input currents are '11'. (d) Input currents going into an AND/NAND device showing the two separate logical '1' input currents (blue, orange) and the total current (black).

Alternatively, a larger device fanout could be achieved by increasing the width of the wire to reduce the MTJ resistance.

We avoid this because this could also change the DW's threshold current and its magnetic shape anisotropy energy. As another alternative, a larger range of device fanouts could be accessed by adding current-limiting resistors to the OUT pin of the device, but this introduces additional area and variability to the circuit.

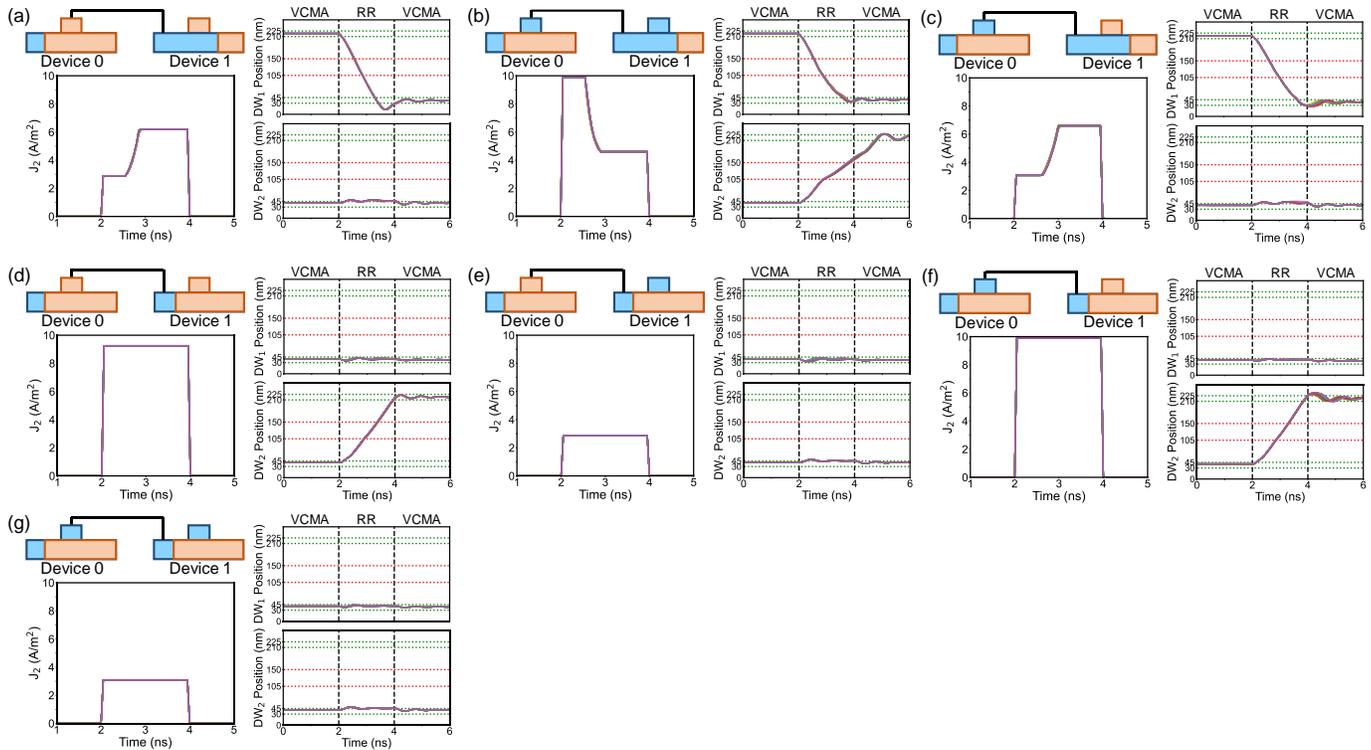

**Figure S4:** Simulation results with the different configurations of the DW-MTJs with TMR = 115%. (a) Device 0's MTJ is P, Device 1's MTJs is AP, and the DW is initially on the right side during a RR pulse. (b) Device 0's MTJ is AP, Device 1's MTJs is P, and the DW is initially on the right side during a RR pulse. (c) Both devices' MTJs are AP and the DW is initially on the right side during a RR pulse. (d) Both devices' MTJs are P and the DW is initially on the left side during a RR pulse. (e) Device 0's MTJ is P, Device 1's MTJs is AP, and the DW is initially on the left side during a RR pulse. (f) Device 0's MTJ is AP, Device 1's MTJs is P, and the DW is initially on the left side during a RR pulse. (g) Both devices' MTJs are AP and the DW is initially on the left side during a RR pulse.

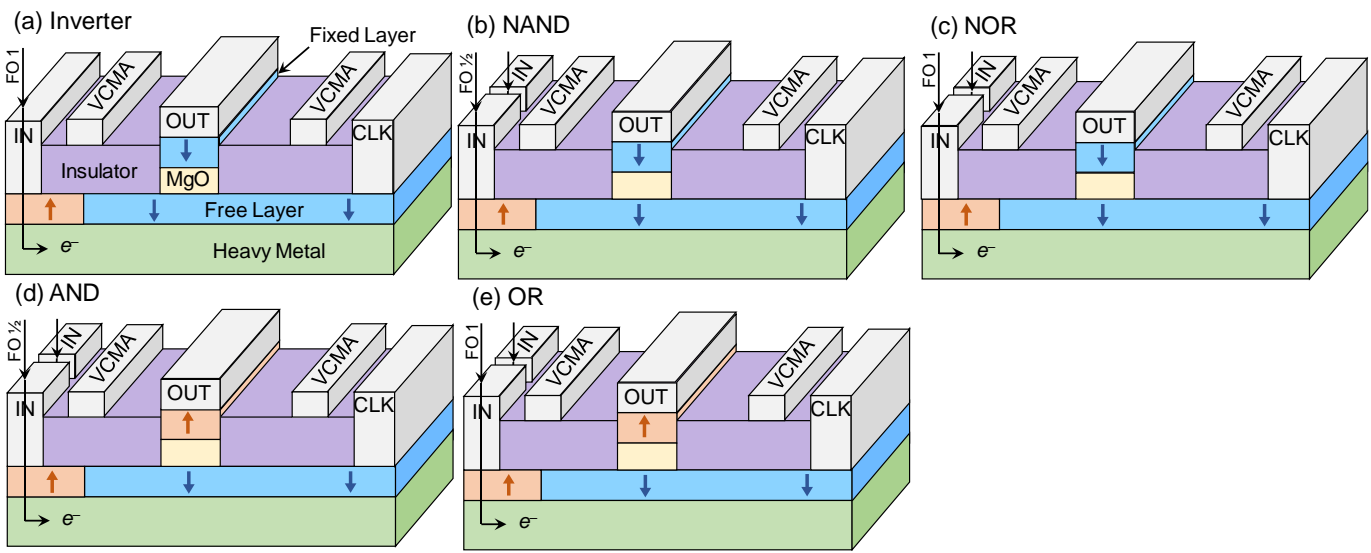

**Figure S5:** DW-MTJ gate schematics (a) DW-MTJ Inverter with the device connected to the IN terminal having a fanout of 1 (b) DW-MTJ NAND with the devices connected to the IN terminals having a fanout of 0.5 (c) DW-MTJ NOR with the devices connected to the IN terminals having a fanout of 1 (d) DW-MTJ AND with the devices connected to the IN terminals having a fanout of 0.5 (e) DW-MTJ OR with the devices connected to the IN terminals having a fanout of 1.